\documentclass[twocolumn,tighten]{aastex62}

\usepackage{soul}

\newcommand{\fracb}[2]{\left(\frac{#1}{#2}\right)}

\newcommand{\mean}[1]{\langle{#1}\rangle}
\newcommand{\red}[1]{\textcolor{black}{#1}}

\graphicspath{{./}{figures/}}

\shorttitle{Top-Hat Jets and the Afterglow of GW170817}
\shortauthors{Gill et al. 2019}

\begin{document}

\title{Numerical Simulations of an Initially Top-Hat Jet and the Afterglow of GW170817$\,$/$\,$GRB170817A}

\author[0000-0003-0516-2968]{Ramandeep Gill}
\affil{Department of Natural Sciences, The Open University of Israel, P.O Box 808, Ra'anana 43537, Israel}
\affil{Physics Department, Ben-Gurion University, PO Box 653, Beer-Sheva 84105, Israel}

\author[0000-0001-8530-8941]{Jonathan Granot}
\affil{Department of Natural Sciences, The Open University of Israel, P.O Box 808, Ra'anana 43537, Israel}
\affil{Department of Physics, The George Washington University, Washington, DC 20052, USA}

\author{Fabio De Colle}
\affil{Instituto de Ciencias Nucleares, Universidad Nacional Aut\'onoma de M\'exico, A. P. 70-543 04510 D. F. Mexico}

\author{Gerardo Urrutia}
\affil{Instituto de Ciencias Nucleares, Universidad Nacional Aut\'onoma de M\'exico, A. P. 70-543 04510 D. F. Mexico}




\begin{abstract}
The afterglow of GRB$\,$170817A/GW$\,$170817 was very unusual, slowly rising as 
$F_\nu\propto{}t_{\rm{}obs}^{0.8}\nu^{-0.6}$, peaking at $t_{\rm{obs,pk}}\sim\,150\;$days, and sharply decaying as 
$\sim{}t_{\rm{}obs}^{-2.2}$. VLBI observations revealed an unresolved radio afterglow image whose flux centroid 
apparently moved superluminally with $v_{\rm{app}}\approx4c$ between $75$ and $230\;$days, 
clearly indicating that the afterglow was dominated by a relativistic jet's compact core. 
Different jet angular structures successfully explained the afterglow lightcurves: Gaussian and steep power-law 
profiles with narrow core angles $\theta_c\lesssim5^\circ$ and significantly larger viewing angles 
$\theta_{\rm{}obs}/\theta_c\sim3-5$. However, a top-hat jet (conical with sharp edges at 
$\theta=\theta_0$) was ruled out since it appeared to produce an early flux rise much steeper 
($\propto{}t_{\rm{}obs}^{a}$ with $a\gtrsim3$) than observed. 
\red{Using 2D relativistic hydrodynamic
simulations of an initially top-hat jet} we show that the initial steep flux rise is an artifact caused by the simulation's finite start time, $t_0$, missing its flux contributions from $t<t_0$ and sometimes ``compensated'' using an analytic 
top-hat jet. \red{While an initially top-hat jet is not very physical, such simulations are particularly useful at $t_{\rm{}obs}\gtrsim{}t_{\rm{obs,pk}}$ when the afterglow emission is dominated by the jet's core and becomes insensitive to its exact initial angular profile if it drops off sharply outside of the core.}
We \red{demonstrate} that an initially top-hat jet \red{fits} GW$\,$170817/GRB$\,$170817A's afterglow lightcurves and flux centroid motion \red{at $t_{\rm{}obs}\gtrsim{}t_{\rm{obs,pk}}$},
for $\theta_{\rm{}obs}/\theta_0\approx3$ and \red{may also fit the earlier lightcurves for} $\Gamma_0=\Gamma(t_0)\gtrsim10^{2.5}$.
\red{We analytically express the degeneracies between the model parameters, and find a minimal jet energy of $E_{\rm{}min}\approx5.3\times10^{48}\;$erg and circum-burst medium density of $n_{\min}\approx5.3\times10^{-6}~{\rm cm}^{-3}$.}

\end{abstract}


\keywords{gamma-ray burst: general ---
ISM: jets and outflows ---
hydrodynamics ---
methods: numerical ---  
relativistic processes --- 
gravitational waves}


\section{Introduction}

The first gravitational wave (GW) detection of a binary neutron star (NS) merger, GW$\,$170817 \citep{Abbott+17-GW170817-Ligo-Detection}, 
was accompanied by the first electromagnetic counterpart to any GW detection -- the weak, short duration gamma-ray burst, 
GRB$\,$170817A \citep{Abbott+17-GW170817-GRB170817A}, that
originated in the nearby ($D\approx40\,$Mpc) elliptical galaxy NGC 4993 \citep{Coulter+17}. An impressive observational campaign detected the 
quasi-thermal kilonova emission
in the NIR-optical-UV energy 
bands over the next few weeks \citep[see, e.g.,][and references therein]{Abbott+17-GW170817A-MMO}. 
The non-thermal afterglow emission was detected after $8.9\,$days in X-rays \citep{Troja+17} and 
after $16.4\,$days in the radio  \citep{Hallinan+17}.

GW$\,$170817/GRB$\,$170817A's long-lived X-ray to radio afterglow emission was highly unusual. In contrast to the 
flux decay seen in almost all GRB afterglows, it showed an exceptionally long-lasting 
flux rise, as $F_\nu(t_{\rm{obs}})\propto\nu^{-0.6}t_{\rm{obs}}^{0.8}$, up to the peak at $t_{\rm{obs,pk}}\sim 150\,$days post merger 
\citep[e.g.][]{Margutti+18,Mooley+18a}, followed by a sharp decay as
$F_\nu\propto{}t_{\rm{obs}}^{a}$ where $a\simeq-2.2$ \citep{Mooley+18b,vanEerten+18}. 
The broadband (X-rays, radio, and 
late-time optical) afterglow emission is consistent with arising from a single power-law 
segment (PLS) of the afterglow synchrotron spectrum, $\nu_m\leq\nu\leq\nu_c$.\footnote{Here $\nu_m$ 
is the synchrotron frequency of minimal energy electrons and 
$\nu_c$ of electrons that cool on the dynamical time \citep{Sari+98}.}


Almost all successful off-axis jet models for this afterglow have an angular profile that is 
either a (quasi-) Gaussian or a narrow core with sharp power-law wings 
\citep{Lamb-Kobayashi-18,Lazzati+18,Troja+17,DAvanzo+18,Gill-Granot-18b,Margutti+18,Resmi+18,Troja+18}.
Moreover, several works have argued that a top-hat jet can be ruled out 
\citep[e.g.,][]{Margutti+18,Mooley+18a} since it would produce a very sharp initial flux rise ($F_\nu\propto{}t_{\rm{obs}}^{a}$ with 
$a\gtrsim3$) compared to the observed one. 
Such a sharp initial flux rise was obtained both 
numerically from 2D hydrodynamic simulations \citep[e.g.,][]{vanEerten-MacFadyen-11,Granot+18b}, and analytically 
assuming an idealized top-hat jet 
\citep[e.g.,][]{Granot+02,Eichler-Granot-06,Nakar-Piran-18}


Here we show that while an idealized top-hat jet would indeed produce sharply rising early lightcurves for off-axis observers, a more realistic description of the dynamics (using numerical 
simulations) for an initially top-hat jet leads to a much shallower flux rise that can explain the GRB$\,$170817A afterglow observations (lightcurves, flux centroid motion, and upper limits on the image size). 
The main difference arises since within the simulation's first dynamical time an 
initial top-hat jet develops a bow-shock like angular structure, which produces afterglow emission resembling 
that from a core-dominated structured jet,\footnote{\red{I.e. a jet in which most of the energy resides within a narrow core, outside of which the energy per solid angle sharply drops.}} with a much shallower flux rise, making the two models 
practically indistinguishable \red{at $t_{\rm{}obs}\gtrsim{}t_{\rm{obs,pk}}$, and not always that easy to distinguish between even at earlier times}.  
%
Numerical simulations have a finite lab-frame start time, $t=t_0>0$, thus missing contributions to $F_\nu$ from $t<t_0$. This is often compensated for by adding emission at $t<t_0$ from a conical wedge from the \citet[][hereafter BM76]{Blandford-McKee-76} spherical self-similar solution \citep[e.g.,][]{vanEerten+12,DeColle+12a,DeColle+12b,Bietenholz+14,Granot+18a,Granot+18b}. This still results in an unphysically sharp flux rise at early observed times, $t_{\rm{obs}}\lesssim2t_{\rm{obs},0}$, corresponding to lab-frame times $t\lesssim2t_0$.

The effects of $t_0$ including $t_{\rm{obs},0}(\theta_{\rm{obs}},t_0)$ are analytically explained in \S~\ref{sec:t_start}.   
The effect of starting the simulations with a larger Lorentz factor (LF) $\Gamma_0=\Gamma(t_0)$ and 
correspondingly smaller $t_0$ is shown in \S~\ref{sec:sim} through 2D relativistic hydrodynamic 
simulations. 
In \S~\ref{sec:scalings} model scalings and the minimal energy and circum-burst medium density estimates are provided. 
In \S~\ref{sec:FC-image-size} we calculate and compare the flux centroid location and the image size and shape 
with radio afterglow measurements of GW$\,$170817/GRB$\,$170817A. Our conclusions are discussed in \S~\ref{sec:dis}.

\section{The Effect of Simulation Start Time}
\label{sec:t_start}
We perform 2D relativistic hydrodynamical simulations with initial conditions of a conical wedge 
of half-opening angle $\theta_0$ taken out of the BM76 solution. This initially narrow and relativistic 
jet expands into a cold circum-burst medium (CBM) with a power-law rest-mass density profile with 
radius $R$ from the central source, $\rho(R)=AR^{-k}$, where for uniform (wind-like) density  environment 
$k=0$ $(k=2)$. The BM76 spherical self-similar phase occurs after the original outflow is significantly 
decelerated 
and most of the energy is in the 
shocked CBM behind the forward (afterglow) shock. The material just behind the shock moves with velocity $\beta c$, 
with $c$ being the speed of light, and bulk LF $\Gamma=(1-\beta^2)^{-1/2}=\Gamma_{\rm{shock}}/\sqrt{2}$. 
The BM76 phase reasonably holds for a top-hat jet while $\Gamma>1/\theta_0$ (assuming $\Gamma_0\theta_0\gg1$, 
as typically inferred for GRBs) before significant lateral spreading can occur. 

The radial width behind the forward shock containing most of the blastwave's energy is $\Delta\sim0.1\,R/\Gamma^2$. 
During the BM76 self-similar phase $\Gamma^2R^{3-k}=\Gamma_0^2R_0^{3-k}=(17-4k)E_{\rm{k,iso}}/16\pi{}Ac^2={\rm{const}}$, 
with $R_0=R(t_0)\approx{}ct_0$ being the initial shock radius. Thus the initial radial width $\Delta_0=\Delta(t_0)\sim0.1R_0/\Gamma_0^2\propto{}R_0^{4-k}\propto\Gamma_0^{-2(4-k)/(3-k)}$ ($\propto\Gamma_0^{-8/3}$ for $k=0$) 
becomes much narrower and harder to resolve for larger $\Gamma_0$ or correspondingly smaller $t_0\approx{}R_0/c\propto\Gamma_0^{-2/(3-k)}$ 
($\propto\Gamma_0^{-2/3}$ for $k=0$). This practically limits $\Gamma_0$ from above and $t_0$ from below. 

An on-axis observer ($\theta_{\rm{}obs}<\theta_0$) receives the first photons from the simulation after a radial time delay of 
\begin{equation}\label{eq:tobs,r}
\frac{t_{\rm{obs},r}}{(1+z)}=t_0-\frac{R_0}{c}\approx\frac{R_0}{4(4-k)c\Gamma_0^2}\approx\frac{t_0}{4(4-k)\Gamma_0^2}~,
\end{equation}
$z$ being the source's cosmological redshift.
For an off-axis observer ($\Delta\theta\equiv\theta_{\rm{}obs}-\theta_0>0$), there is an additional angular time delay,
\begin{eqnarray}\nonumber
\frac{t_{\rm{obs},\theta}}{(1+z)}&=&\frac{R_0}{c}[1-\cos(\Delta\theta)]\approx\frac{\Delta\theta^2}{2}t_0\\~\label{eq:tobs0}
&\approx&\frac{\Delta\theta^2}{2}\left[\frac{(17-4k)E_{\rm{k,iso}}}{16\pi{}Ac^{5-k}\Gamma_0^2}\right]^\frac{1}{3-k}~,
\end{eqnarray}
\citep[e.g.,][]{Granot+17}, which dominates the total time delay 
$t_{\rm{obs},0}=t_{\rm{obs},r}+t_{\rm{obs},\theta}\approx{}t_{\rm{obs},\theta}$ for $\Delta\theta>1/\Gamma_0$. For such off-axis viewing angles one can conveniently express $\Gamma_0\propto{}t_{\rm{obs},0}^{-(3-k)/2}$, which for $k=0$, $E_{\rm{k,iso}}\approx(2/\theta_0^2)E$ and $z\ll1$ gives
\begin{eqnarray}\nonumber
\Gamma_0&\approx&\sqrt{\frac{17E\theta_0^{-2}(\Delta\theta)^6}{64\pi{}nm_pc^{5}t_{\rm{obs},0}^{3}}}\\~\label{eq:Gamma0}
&=&149E_{50.3}^{1/2}n_{-3.6}^{-1/2}\theta_{0,-1}^{-1}\fracb{\Delta\theta}{0.21}^3\fracb{t_{\rm{obs},0}}{10\,\rm{d}}^{-3/2}~,
\end{eqnarray}
where for the numerical value we normalize by our best-fit model parameters derived in \S~\ref{sec:sim}, for which $t_{\rm{obs},0}=38.1,\,23.0,\,18.3\;$days for $\Gamma_0=20,\,40,\,60$.

The compactness argument implies that GRB jets typically have $\Gamma_0\gtrsim100$ for the emission region to be optically thin to $\gamma\gamma$-annihilation \citep[e.g.][]{Lithwick-Sari-01}. 
Such large $\Gamma_0$ are very difficult to simulate, and current numerical works usually set $\Gamma_0\sim20-25$ \citep[see, however,][]{vanEerten-MacFadyen-13}.

\begin{figure}
    \centering
    \includegraphics[width=0.48\textwidth]{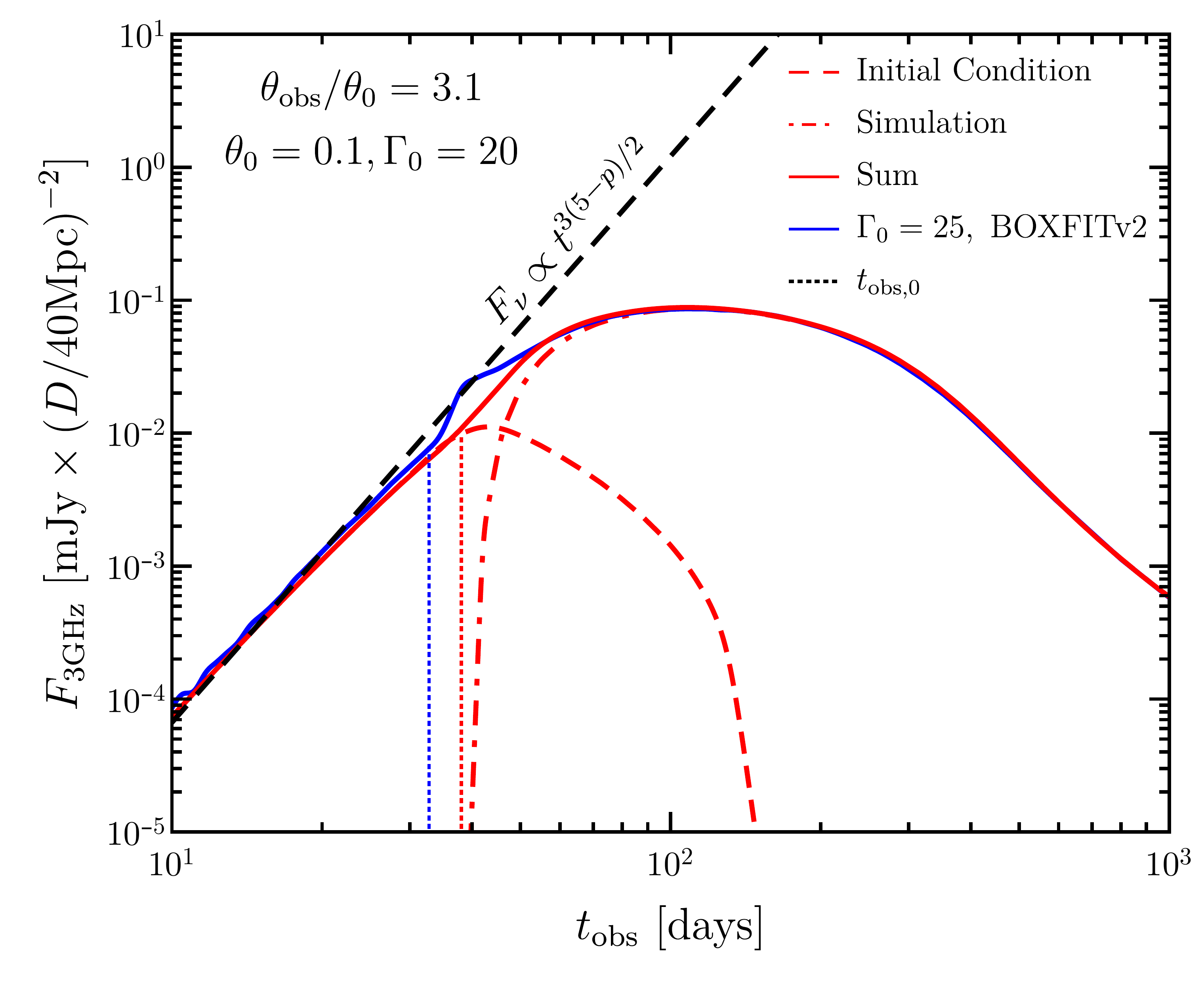}
    \caption{
    Simulated lightcurve decomposition into the synthetic part, obtained from the 
    initial condition (top-hat jet), and that obtained from the simulated region 
    for $t_{\rm{obs}}>t_{\rm{obs,0}}$. Comparison is made with lightcurve from BOXFITv2 code ($\Gamma_0=25$) 
    for the same model parameters (see Fig.~\ref{fig:model-fit}). Extension of both 
    lightcurves at $t_{\rm{obs}}<t_{\rm{obs,0}}$ matches the analytical flux scaling 
    for an off-axis relativistic top-hat jet (the slightly shallower slope towards $t_{\rm{obs,0}}$ arises because of its proximilty to $t_{\rm{obs,pk}}$). }
    \label{fig:components}
\end{figure}

Simulations initialized at $t_0$
do not contribute any flux at $t_{\rm{obs}}<t_{\rm{obs},0}$ (see Fig.~\ref{fig:components}). 
Over the first dynamical time ($t_0<t\lesssim2t_0$), as the simulated jet relaxes from its artificially sharp top-hat initial condition, 
the flux sharply rises at times 
$t_{\rm{obs},0}\leq{}t_{\rm{obs}}\lesssim2t_{\rm{obs},0}$, 
after which the flux evolves smoothly with time. During this relaxation phase, the top-hat jet is slowed down due 
to its interaction with the CBM and develops a bow-shock like structure 
\citep[e.g.][]{Granot+01,vanEerten-MacFadyen-11,DeColle+12b}. 
Its structure at this point resembles a `structured jet' with a highly energetic core, whose velocity is almost radial, surrounded by less energetic 
slower-moving material whose velocity points more sideways. Therefore, an initially top-hat jet inevitably transforms 
into a structured jet. The slower material at angles 
$\theta>\theta_0$ has a much wider beaming cone and its emission starts dominating the off-axis flux. 
As the jet gradually decelerates, its beaming cone widens and off-axis observers start to 
receive flux from smaller $\theta$ closer to the jet's core, resulting in a more gradual flux rise compared 
to an analytic perpetually sharp-edged jet.

To compensate for the missing flux at $t_{\rm{obs}}<t_{\rm{obs},0}$, as shown in Fig.~\ref{fig:components}, 
lightcurves derived from numerical simulations 
are often supplemented with synthetic lightcurves obtained for the initial conditions \citep[usually a conical wedge from the BM76 self-similar solution, e.g.,][]{vanEerten+12,DeColle+12a,DeColle+12b,Bietenholz+14,Granot+18a,Granot+18b}
over a wide range of earlier lab-frame times, $t_*<t<t_0$ with $t_*\ll{}t_0$.
We also compare the lightcurve obtained from the publicly available afterglow modeling 
code BOXFITv2 \citep{vanEerten+12}, which has been widely used to fit afterglow observations of GRB$\,$170817A. 
Lightcurves obtained from our numerical simulations are in excellent agreement with that obtained from BOXFITv2. 
%
%

The observed flux density is given by \citep[e.g.][]{Granot-05,Granot-Ramirez-Ruiz-12}
\begin{equation}\label{eq:Fnu}
F_\nu(t_{\rm obs})=\frac{(1+z)}{4\pi d_L^2(z)}\int{}dt\,\delta_t\int\delta_D^3dL'_{\nu'}\propto\delta_D^3L'_{\nu'}~,
\end{equation}
where $d_L(z)$ is the luminosity distance, the $\delta$-function, 
$\delta_t=\delta\left(t-t_{\rm{obs}}/(1+z)-R\tilde\mu/c\right)$, accounts for the photon arrival 
times \citep{Granot+99}, $R\tilde\mu=\hat{n}\cdot\vec{R}$ where $\hat{n}$ is the direction to the 
observer and $\vec{R}$ is the radius vector (measured from the central source) of each fluid element 
having velocity $\vec v=\vec{\beta}c$ and Doppler factor $\delta_D=[\Gamma(1-\hat{n}\cdot\vec{\beta})]^{-1}$. 
For radial velocities (e.g. a spherical shell), $\hat{n}\cdot\vec{\beta}=\beta\tilde\mu$ and 
$\delta_D\approx2\Gamma/[1+(\Gamma\tilde\theta)^2]$ for $\Gamma\gg1$. 
In Eq.~(\ref{eq:Fnu}), $F_\nu\propto\delta_D^{3}L'_{\nu'}$ holds where $L'_{\nu'}$ and $\delta_D$ are those of the part of the source that dominates the observed emission, which for a top-hat jet viewed off-axis is within an angle $\sim\max(\Gamma^{-1},\Delta\theta)$ of the point in the jet closest to the observer (where $\tilde{\theta}\approx\Delta\theta$), occupying a solid angle $\Omega_*\sim\min[\max(\Gamma^{-2},\Delta\theta^2),\theta_0^2]$. During the early flux-rising phase while the radiation is beamed away from the observer ($\Gamma>1/\Delta\theta$), $\Omega_*={\rm{const}}$ and one can use the scalings of $L'_{\nu'}$ for a spherical flow, 
$L'_{\nu'}\propto{}R^a\nu'^b\propto{}R^a\delta_D^{-b}$, where the PLS-dependent power-law indices $a$ and $b$ are explicitly calculated in \citet{Granot-05}. 
%
%
Therefore, $F_\nu\propto\delta_D^{3-b}R^a$ where 
\citep[e.g.][]{Salmonson-03,Granot-05} $\delta_D\approx2/\Gamma\Delta\theta^2\propto{}R^{(3-k)/2}\Longrightarrow{}F_\nu\propto{}R^{[2a+(3-k)(3-b)]/2}$. 
For GRB$\,$170817A, PLS~G is relevant and $a=[15-9p-2k(3-p)]/4$, $b=(1-p)/2$. From Eq.~(\ref{eq:tobs0}), $t_{\rm{obs}}\propto{}R$ which implies $F_\nu\propto{}t_{\rm{obs}}^{3(5-p)/2}$ for a 
uniform CBM ($k=0$).

In Fig.~\ref{fig:components}, we show the extension of the lightcurve to $t_{\rm{obs}}<t_{\rm{obs},0}$,
where we reproduce the analytic flux scaling derived above. It is clear that BOXFITv2 also supplements the lightcurve at early times ($t<t_0\Leftrightarrow{}t_{\rm{obs}}<t_{\rm{obs},0}$) with the flux from a conical wedge out of the BM76 self-similar solution (also used for the initial conditions).
Although BOXFITv2 allows the user to not include this extension in the final lightcurve, many works 
indeed do include it, even when fitting to observations. 
Either way, the flux at $t_{\rm{obs}}\lesssim2t_{\rm{obs},0}$ is strongly affected by the rather arbitrary simulation start time $t_0$.
Initializing the simulation at a smaller $t_0$ corresponding to a larger $\Gamma_0$ would shift this feature to earlier times and 
recover the much shallower flux rise in the lightcurve.

\begin{figure*}
    \centering
    \includegraphics[width=0.98\textwidth]{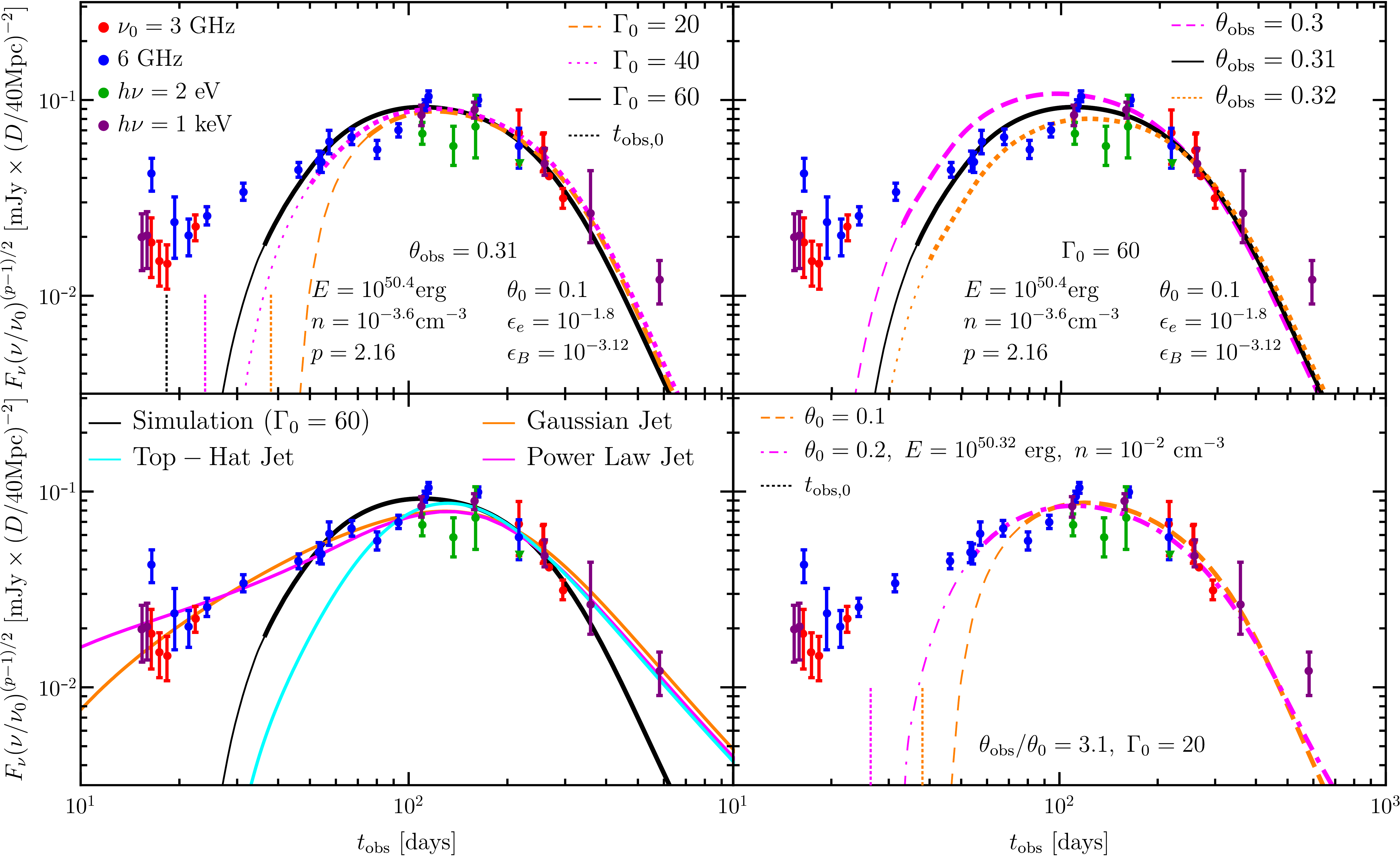}
    \caption{Comparison of simulated afterglow lightcurves for an initially top-hat jet with observations, 
    for different $\Gamma_0$ ({\bf\textit{top-left}}), slightly different viewing angles 
    $\theta_{\rm{obs}}$ ({\bf\textit{top-right}}), different $\theta_0$ ({\bf\textit{bottom-right}}), 
    and \red{semi-analytic models of different jet structures ({\bf\textit{bottom-left}}; see text for model parameters)}. 
    Observations in different energy bands (with late-time X-ray data from \citealt{Haggard+2018,Hajela+2019}) are normalized to the corresponding flux density 
    at $\nu_0=3$~GHz. Upper limits are marked by downward triangles. The simulated-flux deficiency at 
    $t_{\rm{obs}}\lesssim2t_{\rm obs,0}$ is an artefact of starting the simulation with low $\Gamma_0$ and 
    at a correspondingly large lab-frame time $t_0$. No simulation flux is available at 
    $t<t_0\Leftrightarrow t_{\rm{obs}}<t_{\rm{obs},0}$. 
    }
    \label{fig:model-fit}
\end{figure*}

\section{Different $\Gamma_0$ fits to the Afterglow Data of GW$\,$170817/GRB$\,$170817A}
\label{sec:sim}

Here we show results of 2D hydrodynamic simulations using the special-relativistic 
hydrodynamics code \emph{Mezcal}, post-processed by a complimentary radiation code 
\citep[see][for details]{DeColle+12a,DeColle+12b}. The simulations are initialized with a conical wedge of half-opening angle $\theta_0=0.1,\,0.2\;$rad and initial LF $\Gamma_0=20,\,40,\,60$ expanding into a uniform CBM ($k=0$) of rest-mass  density 
$\rho_0=nm_p$ and number density $n$, $m_p$ being the proton mass. 
The outflow has an isotropic-equivalent kinetic energy $E_{\rm{k,iso}}=10^{53}\;$erg, 
corresponding to a true jet energy of $E=(1-\cos\theta_0)E_{\rm{k,iso}}\approx5\times10^{50}\;$erg for $\theta_0=0.1$ 
and $E\approx2\times10^{51}\;$erg for $\theta_0=0.2$.
%

We consider synchrotron radiation from relativistic electrons that are accelerated at the 
afterglow shock to a power-law energy distribution, $dN_e/d\gamma_e\propto\gamma_e^{-p}$
for $\gamma_e>\gamma_m$ with $p=2.16$, which are a fraction $\xi_e$ of all post-shock electrons, and hold a fraction $\epsilon_e=0.1$ of the post-shock internal energy density, where a fraction $\epsilon_B=0.1$ goes to the magnetic field.
%
The radiation is calculated numerically for a fixed set of model parameters ($E,\,n,\,\epsilon_e,\,\epsilon_B,\,p,\,\theta_0$) 
and for a grid of $\theta_{\rm{obs}}$ values. \red{When including the parameter $\xi_e$, the set of model parameters become degenerate, 
where the afterglow flux is invariant under the change $E\to E/\xi_e$, $n\to{}n/\xi_e$, $\epsilon_e\to\epsilon_e\xi_e$, 
and $\epsilon_B\to\epsilon_B\xi_e$, for $m_e/m_p<\xi_e\leq1$.} We then use the scaling relations described in \citet{Granot-12} for arbitrary values of ($E,\,n$), as well as the scaling with the shock microphysical parameters in each PLS 
 \citep[Table~2 of][]{Granot-Sari-02}. See \citet{Granot+17} for further details.
 
\red{There are in total 8 model parameters, 
i.e. $E,\,n,\,\epsilon_e,$ $\epsilon_B,\,p,\,\xi_e,\,\theta_0,\,\theta_{\rm obs}$.
There are 5 effective observational constraints: (i) 
the spectral index $b\approx-0.58$ 
($F_\nu\propto\nu^b$; $b=[1-p]/2$ for PLS G, which determines $p=1-2b\approx2.16$), (ii)
the lightcurve peak time $t_{\rm obs,pk}\approx150\,$days, (iii) the peak flux $F_{\nu,\rm pk}$, (iv) the shape of the lightcurve near the peak (which approximately determines $\theta_{\rm{}obs}/\theta_0$), (v) the radio flux centroid's apparent velocity. These 5 constraints involve equalities and reduce the dimensionality of the allowed parameter space from an 8D to a 3D. 
There are also 3 additional constraints that involve inequalities 
and hence only reduce its volume but not its dimensionality:
the fact that all the broadband afterglow observations lie within PLS G, $\nu_m<\nu<\nu_c$, and $\theta_{\rm obs}\lesssim0.5$ 
from the GW detection. 
}

Our afterglow lightcurve fitting is guided by the measured peak at
$t_{\rm{obs,pk}}\sim150\;$days \citep{Dobie+18} and the data points near the peak. Fig.~\ref{fig:model-fit} shows 
the fit to the afterglow data 
for different initial $\Gamma_0$ ({\it{}top-left panel}) and viewing angles $\theta_{\rm{obs}}$ ({\it{}top-right panel}). 
We do not attempt to fit the early time data at $t_{\rm{obs}}\lesssim40$~days, before the simulated lightcurves contain the 
dominant and dynamically relaxed contribution from the hydrodynamic simulation. Nevertheless, 
we obtain a reasonable fit to the afterglow data for different values of $\Gamma_0$, where our lightcurves for larger 
$\Gamma_0$ extend to earlier times and can adequately explain the data at $t_{\rm{obs}}\gtrsim40\,$days. 

The best constrained parameters are \citep[also see][]{Granot+18a}: (i) $p\approx2.16$,  
and (ii) $\theta_{\rm{obs}}/\theta_0\approx3.1\pm0.1$, since it significantly affects the shape of the 
lightcurve before and around 
the peak time. In the bottom-right panel of Figure~\ref{fig:model-fit}, we compare the model lightcurves for $\theta_0=0.1,\,0.2$ and show 
that in both cases $\theta_{\rm{obs}}/\theta_0=3.1$ \red{provides a comparably good fit}, while fixing the same values for the shock microphysical 
parameters but  varying the true jet energy $E$ and CBM density $n$. 

\red{We compare the simulation lightcurves with those obtained from semi-analytic models of different jet 
structures, namely a top-hat (THJ), Gaussian (GJ), and a power law jet (PLJ) \citep[see][for models of structured jets]{Gill-Granot-18b}. 
For the top-hat jet we prescribe the same dynamics as that for the two structured jets, i.e. every part of the jet evolves 
locally as if it were part of a spherical flow, with no sideways spreading. As a result, all three semi-analytic models yield very 
similar lightcurves right after the peak when the compact core of the jet becomes visible to the off-axis observer.
On the other hand, the simplified dynamics of the semi-analytic models leads to a significantly shallower post-peak flux decay rate 
compared to the simulated one, which may be attributed to the combination of a shallower asymptotic decay and a smaller overshoot 
just after the peak \citep[e.g.][]{Granot-07}. The post-peak flux decay behavior of different structured jets will be investigated 
in more detail using 2D numerical simulations in another work (Urrutia et al. 2019, in preparation). 
For the semi-analytic models one set of model parameter values that can explain the observations sufficiently well are: 
$E_{\rm k,iso,\{c, jet\}}\approx10^{51.6}\,$erg, $\theta_{\rm \{c,jet\}}\approx5^\circ$, 
$\theta_{\rm obs}=27^\circ$, $\epsilon_e\approx10^{-1}$, $\epsilon_B\approx10^{-2.8}$, and the only difference is in the 
core Lorentz factors between the three models, with $\Gamma_{\rm c}^{\rm PLJ}=100$, $\Gamma_{\rm jet}^{\rm THJ}=\Gamma_{\rm c}^{\rm GJ}=600$.
}

\section{Flux scalings, model degeneracies, and minimum jet energy and CBM density estimates}\label{sec:scalings}
For the lightcurve fits we assume $\xi_e=1$,
and use the dependence on the shock microphysical parameters in PLS~G 
from \citet{Granot-Sari-02}, now including the degeneracy due to $\xi_e$ \citep[e.g.][]{vanEerten-MacFadyen-12}, $F_{\nu,G}\propto\epsilon_e^{p-1}\epsilon_B^{(p+1)/4}\xi_e^{2-p}\nu^{(1-p)/2}$. We also use the global scaling 
relations \citep{Granot-12}, which are conveniently parameterized through 
length and time, 
\begin{equation}\label{eq:scaling}
\alpha=\frac{\ell'}{\ell}=\frac{t'}{t}=\frac{t'_{\rm{obs}}}{t_{\rm{obs}}}=\left(\frac{E'/E}{n'/n}\right)^{1/3}~, 
\end{equation}
and through mass and energy, $\zeta=m'/m=E'/E$, where the rescaled parameters 
are denoted with a prime, $\mathcal F = F'_{\nu,G}(t'_{\rm{obs}},\epsilon'_e,\epsilon'_B,\xi'_e)/F_{\nu,G}(t_{\rm{obs}},\epsilon_e,\epsilon_B,\xi_e)$, 
\begin{equation}\label{eq:Fscaling}
\mathcal F = \zeta^{(p+5)\over4}\alpha^{-3(p+1)\over4}\fracb{\epsilon'_e}{\epsilon_e}^{p-1}
\fracb{\epsilon'_B}{\epsilon_B}^{(p+1)\over4}\fracb{\xi'_e}{\xi_e}^{2-p}~.
\end{equation}
%
\begin{figure}
    \centering
    \includegraphics[width=0.48\textwidth]{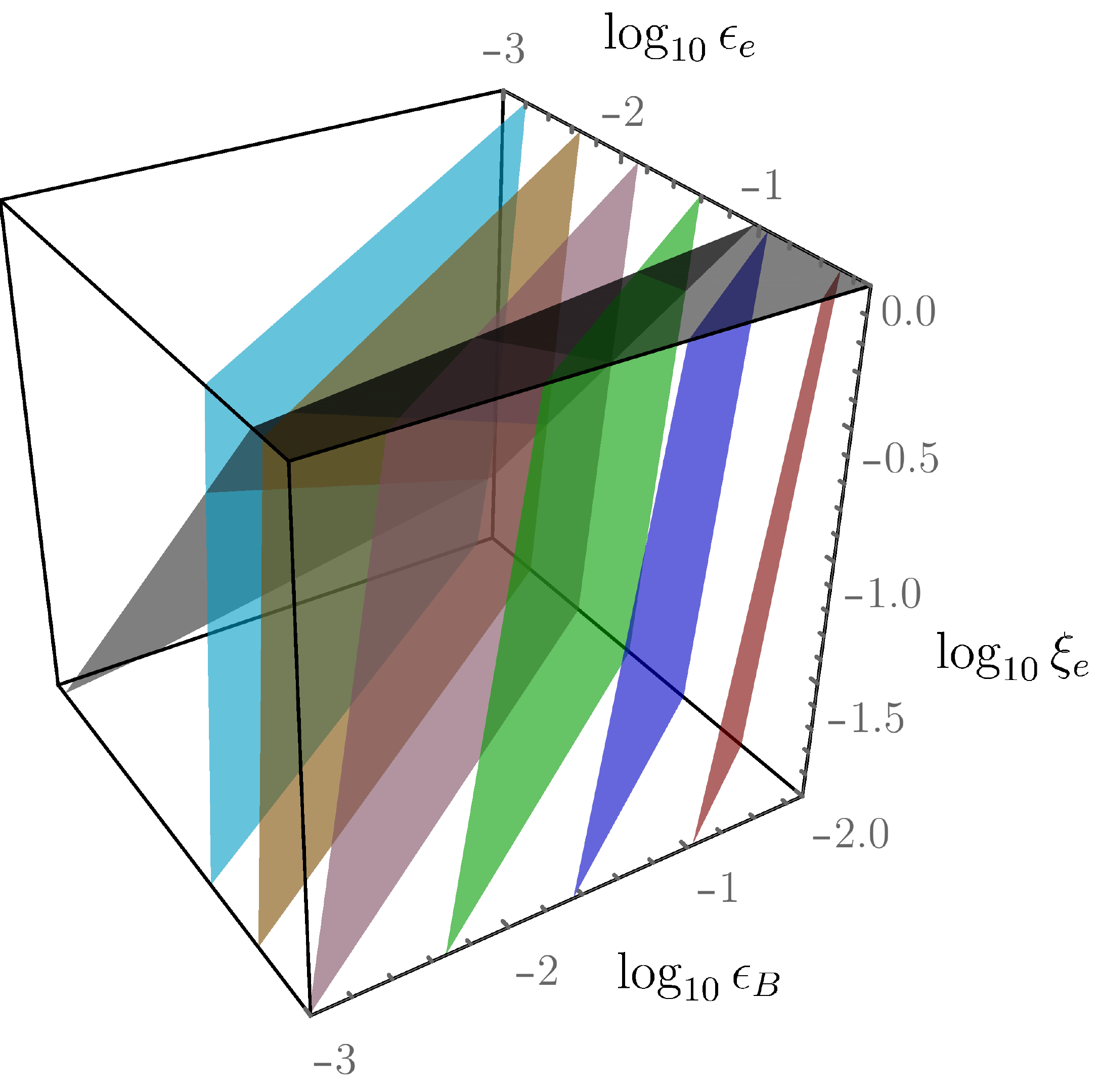}
    \caption{Allowed 3D parameter space [$\xi_e,\epsilon_e,\epsilon_B$] shown by planes 
    in this space for different jet energies, $\log_{10}E = 48.3, 48.6,...,49.8$ (from red to cyan) 
    following Eq.~(\ref{eq:scaling2a}), which is satisfied in the region above the black plane for which 
    $\xi_{e,\min}\leq\xi_e\leq1$. The constraint on $\xi_{e,\min}$ from Eq.~(\ref{eq:xie-min}) is shown by 
    the black plane. The excluded region, for which $\xi_{e,\min}>1$, is shown by the shaded transparent 
    region on the top-face of the cube.}
    \label{fig:intersections}
\end{figure}
%
Next, we constrain $E$ from below by using these scaling relations and our (partly degenerate) 
best-fit parameters: $E=10^{50.4}\;{\rm{erg}}$, $n=10^{-3.6}\;{\rm{}cm^{-3}}$, $\epsilon_e=10^{-1.8}$, 
$\epsilon_B=10^{-3.12}$, $\theta_{\rm{obs}}/\theta_0=3.1$ (fixing $\xi_e=1$, $p=2.16$, $\theta_0=0.1$). Matching the peak time of 
the simulated lightcurve to $t_{\rm{obs,pk}}\approx150\,$days requires no significant time rescaling, and yields 
$\alpha=t'_{\rm{obs}}/t_{\rm{obs}}\approx1$. Matching the peak flux to that observed requires equating Eq.~(\ref{eq:Fscaling}) to 
unity. Altogether, replacing the unprimed quantities by the best-fit values, and then making the rescaled quantities unprimed, 
and solving for $\zeta$, yields
\begin{eqnarray}\label{eq:scaling2a}
\zeta&&=\frac{E}{10^{50.4}\,{\rm{erg}}}=\frac{n}{10^{-3.6}\,{\rm{cm}}^{-3}} \nonumber \\
\label{eq:scaling2b}
&&\approx\fracb{\epsilon_e}{10^{-1.8}}^{4(1-p)\over(p+5)}\fracb{\epsilon_B}{10^{-3.12}}^{-(p+1)\over(p+5)}\xi_e^{4(p-2)\over(p+5)}~,
\end{eqnarray}
where the equality in Eq.~(\ref{eq:scaling2a}) results from Eq.~(\ref{eq:scaling}) when $\alpha=1$. 
This leaves us with a 3D allowed parameter space since we started with 7 free model parameters ($\theta_0=0.1$ was fixed by 
the simulation, leaving $E,\,n,\,\epsilon_e,$ $\epsilon_B,\,p,\,\xi_e,\,\theta_{\rm obs}$) and used 4 observational constraints. 
The jet energy in Eq.~(\ref{eq:scaling2a}) 
decreases with increasing $\epsilon_e,\,\epsilon_B$ and increases only weakly with $\xi_e$. A minimal energy 
constraint can be obtained by maximizing the values of $\epsilon_e,\,\epsilon_B$ and minimizing that of $\xi_e$. This 
is demonstrated in Fig.~\ref{fig:intersections}, where we show planes in the 3D parameter space [$\xi_e, \epsilon_e, \epsilon_B$] 
for different jet energies. Here we first use the fact that the broadband afterglow observations lie on a single PLS, 
with $\nu>\nu_m$, where we obtain
\begin{equation}\label{eq:nu_m}
    \nu_m = 8.93\times10^5~\xi_e^{-2}E_{{\rm k,iso},52.7}^{1/2}
    \epsilon_{e,-1.8}^2\epsilon_{B,-3.12}^{1/2}t_{\rm obs,150d}^{-3/2}\,{\rm Hz}
\end{equation}
for $t_{\rm obs,150d}=t_{\rm obs}/(150\,{\rm days})$ and $p=2.16$ from the expression for PLS G given in \citet{Granot-Sari-02}. 
This expression is only valid for a spherical flow and for an on-axis observer, for 
whom the flux is dominated by emission from material along the LOS. At $t_{\rm obs}\geq t_{\rm obs,pk}\approx150\,$days, the flux is dominated 
by that from the core of the jet with $E_{\rm k,iso,c}\lesssim10^{52.7}\,$erg. At $t_{\rm obs}<t_{\rm obs,pk}$, the flux is dominated by emission from 
material outside of the core at $\theta>\theta_0$ with $E_{\rm k,iso}<E_{\rm k,iso,c}$. To obtain the value of $\nu_m$ for an off-axis observer, we 
calibrated Eq.~(\ref{eq:nu_m}) by comparing it with the value of $\nu_m$ obtained from our numerical simulation around the time of the earliest radio observations at $t_{\rm obs}\approx16.4\,$days. 
Next, we use the relation from Eq.~(\ref{eq:scaling2b}) in Eq.~(\ref{eq:nu_m}) and replace $E_{\rm k,iso}$ to obtain an expression that depends 
only on shock microphysical parameters, which, for $\nu_m({\rm 16.4~days})<\nu_{\rm obs}=3\,$GHz, yields a lower limit on $\xi_e$
\begin{equation}\label{eq:xie-min}
    \xi_e>\xi_{e,\min} \approx 0.84~\epsilon_{e,-1}^{6/7}\epsilon_{B,-1}^{1/7}~.
\end{equation}
This constraint is shown as a shaded black plane in Fig.~\ref{fig:intersections} above which Eq.~(\ref{eq:scaling2a}) is satisfied. 
Another useful constraint here is that $\xi_{e,\min}<1$, which yields
\begin{equation}\label{eq:epsB-max}
\epsilon_e<\epsilon_{e,\max}=0.12\epsilon_{B,-1}^{-1/6}\ .
\end{equation}
We first use the constraint on $\xi_{e}$ from Eq.~(\ref{eq:xie-min}) in Eq.~(\ref{eq:scaling2a}) and remove the dependence on $\xi_e$. 
Next, we use the additional constraint on $\epsilon_e$ from Eq.~(\ref{eq:epsB-max}) (which is equivalent to substituting $\xi_e=1$ and $\epsilon_e=\epsilon_{e,\max}$ in Eq.~[\ref{eq:scaling2a}]) to obtain
\begin{equation}\label{eq:Emin}
E_{\min}\approx7.7\times10^{48}\,\epsilon_{B,-1}^{-1/3}\;\rm{erg}=5.3\times10^{48}\epsilon_{B,-0.5}^{-1/3}\;\rm{erg}~,
\end{equation}
as also demonstrated in Fig.~\ref{fig:intersections} by the intersection of the black plane with planes marked by jet energies $E>E_{\min}$.

If we consider only some $\xi_e<1$, as may be expected on theoretical grounds, then Eq.~(\ref{eq:xie-min}) will lead to 
$\epsilon_{e,\max}=0.12\epsilon_{B,-1}^{-1/6}\xi_e^{7/6}$ and accordingly increase $E_{\min}$ to
\begin{eqnarray}\nonumber
E_{\min}&\approx&3.6\times10^{49}\,\epsilon_{B,-1}^{-1/3}\xi_{e,-1}^{-2/3}\;\rm{erg}\\ \label{eq:Emin2}
&=&5.3\times10^{48}\epsilon_{B,-0.5}^{-1/3}\xi_e^{-2/3}\;\rm{erg}~.
\end{eqnarray}
Finally, according to Eq.~(\ref{eq:scaling2b}) $E_{\min}$ also corresponds to a minimal CBM density,
\begin{eqnarray}\nonumber
n_{\min}&\approx&3.6\times10^{-5}\,\epsilon_{B,-1}^{-1/3}\xi_{e,-1}^{-2/3}\;\rm{cm^{-3}}\\ \label{eq:n_min}
&=&5.3\times10^{-6}\epsilon_{B,-0.5}^{-1/3}\xi_e^{-2/3}\;\rm{cm^{-3}}~.
\end{eqnarray}


\section{Model comparison with afterglow image size and flux centroid motion}\label{sec:FC-image-size}
We compare the afterglow image size and flux centroid motion
on the plane of the sky as obtained 
from our simulations to the  GW$\,$170817/GRB$\,$170817A radio observations. VLBI observations between 75 and $230\,$days revealed an unresolved 
source whose flux centroid showed apparent superluminal motion with $\langle v_{\rm{app}}\rangle/c=\langle\beta_{\rm{app}}\rangle=4.1\pm0.5$ \citep{Mooley+18b}. The flux centroid's location on the plane of the sky is defined as 
\begin{equation}
\mathbf{\tilde{r}}_{\rm{fc}}=(\tilde{x}_{\rm{fc}},\,\tilde{y}_{\rm{fc}})=\frac{\int{}dF_\nu\,\mathbf{\tilde{r}}}{\int{}dF_\nu}=\frac{\int{}d\tilde{x}\,d\tilde{y}\,I_\nu\mathbf{\tilde{r}}}{\int{}d\tilde{x}\,d\tilde{y}\,I_\nu}
\end{equation}
\citep[e.g.,][]{Granot+18b}, where $dF_\nu=I_\nu d\Omega=I_\nu{}d_A^{-2}dS_\perp$, with $I_\nu$ being the specific intensity, $d_A$ the angular distance, and $dS_\perp=d\tilde{x}\,d\tilde{y}$ a transverse area element on the plane of the sky. The jet symmetry axis is in the $\tilde{x}$-$\tilde{z}$ plane, where the $\tilde{z}$-axis points to the observer.
Because of the flow's axisymmetry, the image has the reflection symmetry $I_\nu(\tilde{x},\tilde{y})=I_\nu(\tilde{x},-\tilde{y})$. 
Therefore, $\mathbf{\tilde r}_{\rm fc}=(\tilde{x}_{\rm{fc}},0)$ and the flux centroid moves along the $\tilde x$-axis. Since $I_\nu=d_A^2dF_\nu/dS_\perp\propto{}F_\nu/S_\perp$ where $S_\perp\propto\ell^2$, it scales in PLS~G as 
$\mathcal I = I'_{\nu,G}(t'_{\rm{obs}},\tilde{x}',\tilde{y}')/I_{\nu,G}(t_{\rm{obs}},\tilde{x},\tilde{y})$, 
\begin{equation}\label{eq:Inu_scaling}
\mathcal I 
=\zeta^{(p+5)\over4}\alpha^{-(3p+11)\over4}\fracb{\epsilon'_e}{\epsilon_e}^{p-1}\fracb{\epsilon'_B}{\epsilon_B}^\frac{p+1}{4}\fracb{\xi'_e}{\xi_e}^{2-p}~.
\end{equation}
The image size, flux centroid location, and observed time all scale as $\alpha=\tilde{x}'/\tilde{x}=\tilde{y}'/\tilde{y}=\tilde{x}'_{\rm{fc}}/\tilde{x}_{\rm{fc}}=t'_{\rm{obs}}/t_{\rm{obs}}$, independent of the r.h.s of Eq.~(\ref{eq:Inu_scaling}). The flux centroid's apparent velocity $\beta_{\rm{app}}$ remains unchanged, but shifts to the rescaled observer time
\citep[see, e.g. Sec. 4 of][for more details]{Granot+18b}. 

\begin{figure}
    \centering
    \includegraphics[width=0.47\textwidth]{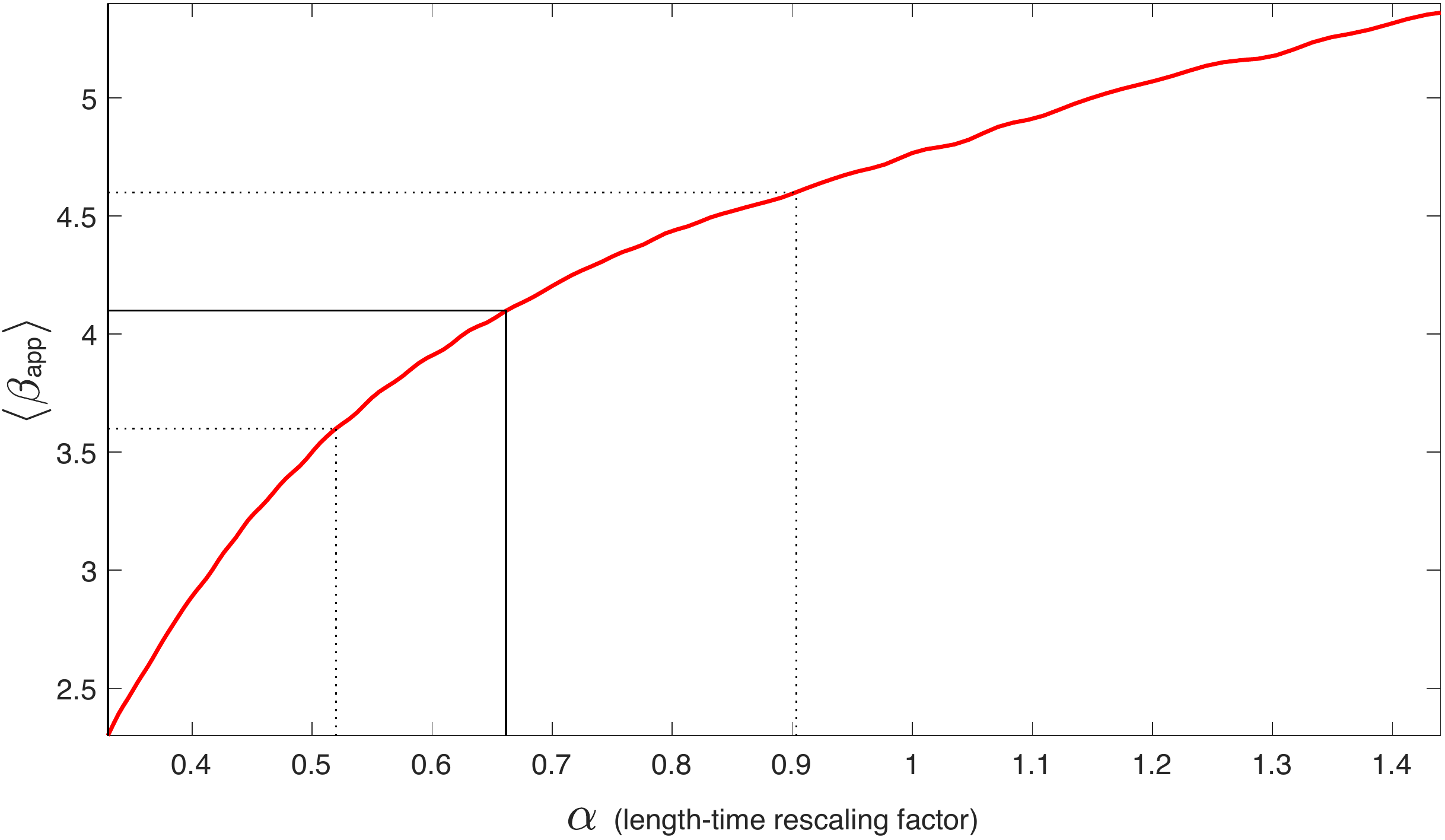}
    \caption{The observed mean radio flux centroid velocity between 75 and 230 days,
    $\mean{\beta_{\rm{app}}}=4.1\pm0.5$ \citep[\textit{horizontal lines};][]{Mooley+18b}, is compared to that from our best-fit simulation (\textit{thick red line}) as a function of $\alpha$. It corresponds to $\alpha=0.661^{+0.242}_{-0.141}$ (\textit{vertical lines}) or a $1\sigma$ confidence interval $0.520 <\alpha<0.903$. 
    }
    \label{fig:alpha-fit}
\end{figure}

\begin{figure}
    \centering
    \includegraphics[width=0.47\textwidth]{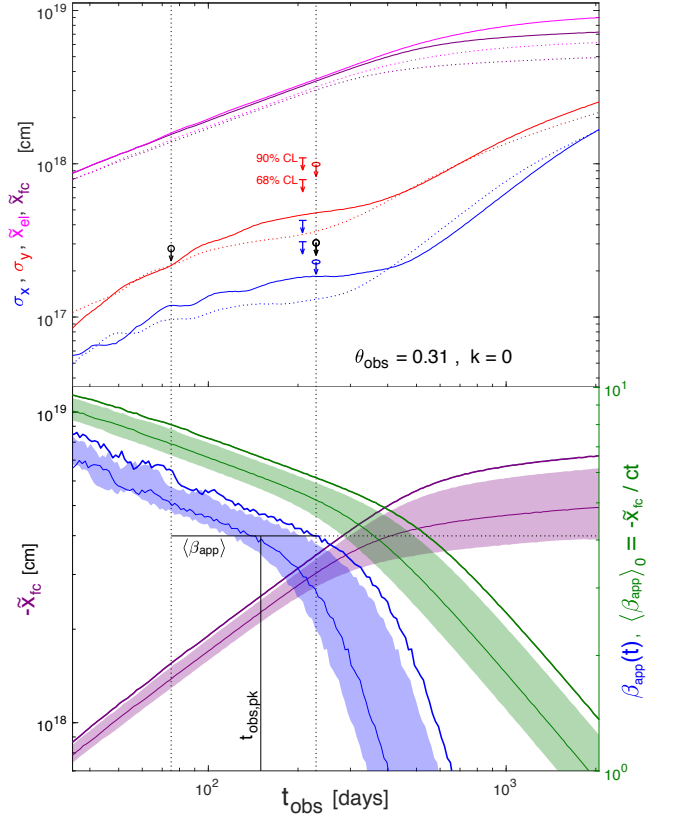}
    \caption{{\bf\textit{Top}}: The evolution of the afterglow image flux-centroid location ($\tilde{x}_{\rm{fc}}$; \textit{deep purple}), and best-fit parameters to an elliptical Gaussian: semi-minor axis $\sigma_x$ (\textit{blue}), semi-major axis $\sigma_y$ (\textit{red}), and center $\tilde{x}_{\rm{el}}$ (\textit{magenta}). 
    Solid lines are for our fiducial model, and dotted lines of the same color are for our 
    best-fit length-time rescaling parameter $\alpha=0.661$. 
    Our model calculations are compared to observational 
    upper limits on the semi-major (\textit{red}) and semi-minor (\textit{blue}). The limits at $75,\,230\;$days \citep{Mooley+18b} are $\sim1\sigma$; ellipse symbols assume a 4:1 axis ratio; black-circle symbols assume a circular Gaussian and apply to both axes. At $207\;$days \citep{Ghirlanda+18} we show $68\%$~CL and $90\%$~CL limits for our calculated axis ratio ($\sigma_y/\sigma_x=2.55$). 
    The vertical dotted black lines indicate the two epochs (75 and 230~days) between which 
    $\mean{\beta_{\rm{app}}}=4.1\pm0.5$ was measured \citep{Mooley+18b}. 
    {\bf\textit{Bottom}}: The 
    evolution of the flux-centroid location (\textit{left $y$-axis}) for our fiducial model 
    (\textit{deep purple}) and its rescaled version to best fit the measured $\mean{\beta_{\rm{app}}}$ (shaded region of matching color for the $1\sigma$ confidence region), as well as of the flux centroid's apparent velocity (\textit{right $y$-axis}). For the latter we show both the 
    mean apparent velocity from $t=0$, $\mean{\beta_{\rm{app}}}_0=\vert\tilde{x}_{\rm{fc}}\vert/ct_{\rm{obs}}$ 
    (\textit{dark green}), and for the instantaneous $\beta_{\rm{app}}=\vert{}d\tilde{x}_{\rm{fc}}/d(ct_{\rm{obs}})\vert$ (\textit{blue}).}
    \label{fig:gaussian-fit}
\end{figure}

Fig.~\ref{fig:alpha-fit} shows how our best-fit simulated $\langle\beta_{\rm{app}}\rangle$ varies with $\alpha$. The measured $\langle\beta_{\rm{app}}\rangle=4.1\pm0.5$ corresponds to $\alpha=0.661^{+0.242}_{-0.141}$, and is consistent (at the 1.35$\sigma$ level) with 
our fiducial model that fits the afterglow lightcurve ($\alpha=1$), which thus passes an important consistency check.

To calculate the afterglow image size and shape, we fit the surface brightness to an elliptical Gaussian,  $I_\nu\propto\exp[-(\tilde{x}-\tilde{x}_{\rm{el}})^2/2\sigma_x^2-\tilde{y}^2/2\sigma_y^2]$ centered at $(\tilde{x}_{\rm{el}},0)$, where $(\sigma_x,\sigma_y)$ are the 
standard deviations of the semi-minor and semi-major axes along the $\tilde{x}$-axis and $\tilde{y}$-axis, respectively \citep{Granot+18b}. 
The top-panel of Fig.~\ref{fig:gaussian-fit} shows the evolution of the afterglow flux-centroid location, and the afterglow image size and shape for $\alpha=1$ and for the $\mean{\beta_{\rm{app}}}$ best-fit $\alpha=0.661$. Our image size is consistent with the upper limits from radio VLBI observations
\citep{Mooley+18b,Ghirlanda+18}. The bottom-panel of Fig.~\ref{fig:gaussian-fit} shows the flux centroid's location, $\tilde{x}_{\rm{fc}}(t_{\rm{obs}})$, 
as well as its instantaneous ($\beta_{\rm{app}}=\vert{}d\tilde{x}_{\rm{fc}}/d(ct_{\rm{obs}})\vert$) and mean 
($\mean{\beta_{\rm{app}}}_0=\vert\tilde{x}_{\rm{fc}}\vert/ct_{\rm{obs}}$) apparent velocities, for our fiducial model ($\alpha=1$), 
and over the $1\sigma$ confidence interval of $\alpha$ derived in Fig.~\ref{fig:alpha-fit}. We find that $\beta_{\rm{app}}(t_{\rm{obs,pk}})\approx\mean{\beta_{\rm{app}}}$.

The measured $\mean{\beta_{\rm{app}}}$ favors a slightly larger $\theta_0$ compared to our $\theta_0=0.1$.
The lightcurve peak occurs when $1/\Delta\theta\approx\Gamma(t_{\rm{obs,pk}})\approx\beta_{\rm{app}}(t_{\rm{obs,pk}})\approx\mean{\beta_{\rm{app}}}$, 
implying $\theta_0\approx[\mean{\beta_{\rm{app}}}(\theta_{\rm{obs}}/\theta_0-1)]^{-1}\approx0.116^{+0.016}_{-0.013}$ 
using the measured $\langle\beta_{\rm{app}}\rangle=4.1\pm0.5$ and our inferred $\theta_{\rm{obs}}/\theta_0=3.1\pm0.1$. 
The latter implies $\Gamma(t_{\rm{obs,pk}})\propto\theta_0^{-1}$, which in turn 
for the measured $t_{\rm{obs,pk}}(\theta_0)\approx150\,{\rm{days}}$, and either pre- or post-jet break simple 
analytic dynamics, implies $E/n\propto\theta_0^{-6}$. This agrees with the best-fit values for our 
$\theta_0=0.1,\,0.2$ to within 34\%, $(0.2/0.1)^6(10^{50.32}/10^{-2})/(10^{50.4}/10^{-3.6})\approx1.337$.
Even for $\theta_0=0.2$, a derivation of $E_{\rm{min}}$ following the one done above for $\theta_0=0.1$ gives 
a result very similar to Eq.~(\ref{eq:Emin}), implying that it is quite robust.
Altogether, $\mean{\beta_{\rm{app}}}$ provides an additional observational constraint that allows us to 
constrain an additional model parameter, $\theta_0$, which still leaves us with a 3D allowed parameter space.

\section{Discussion and Conclusions}
\label{sec:dis}
 This work demonstrates using afterglow lightcurves and image size, shape and flux centroid motion, 
 all derived from 2D hydrodynamical numerical simulations, that an initially top-hat jet can fit the afterglow 
 observations of GW$\,$170817/GW$\,$170817A for $\theta_0\approx0.1$ and $\theta_{\rm{obs}}/\theta_0\approx3$ 
 at $t_{\rm obs}\gtrsim t_{\rm obs,pk}$. 
 We show that simulations of initially top-hat jets with a modest $\Gamma_0\sim20-25$ can only be used to fit 
 the late-time observations near the lightcurve's peak at $t_{\rm{obs,pk}}\approx150\,$days. Fitting earlier 
 observations at $t_{\rm{obs}}\lesssim60\,$days requires $\Gamma_0\gtrsim25$. 

We analytically express the allowed parameter space (Eqs.~[\ref{eq:scaling2b}]) showing the full degeneracies 
between the model parameters, and find a robust lower limit on the jet's true energy, 
$E_{\min}\approx5.3\times10^{48}\,$erg (Eq.~[\ref{eq:Emin}]), and the CBM density, 
$n_{\min}\approx5.3\times10^{-6}~{\rm cm}^{-3}$ (Eq.~[\ref{eq:n_min}]).
 
Our numerical simulations are initialized using a conical wedge from the BM76 self-similar  solution; 
 a similar setup is used in the BOXFITv2 code. The simulation is initialized at a finite lab-frame time 
 $t_0=t(\Gamma_0)$ corresponding to the modest $\Gamma_0=\Gamma(t_0)$. Therefore, no flux contributions 
 are obtained from the simulated region at $t<t_0\Leftrightarrow t_{\rm{obs}}<t_{\rm{obs,0}}$. 
Artificially supplementing the lightcurve at those times with flux arising from the initial 
condition (a top-hat jet) over a wide time-range produces an early sharply-rising flux for an off-axis 
($\theta_{\rm{obs}}>\theta_0$) observer.
%
However, within a dynamical time ($t_0<t\lesssim2t_0\Leftrightarrow{}t_{\rm{obs,0}}<t_{\rm{obs}}\lesssim2t_{\rm{obs,0}}$), as 
the outflow relaxes from the initial conditions it develops a bow-shock like angular structure that resembles a structured jet 
 having an energetic relativistic core surrounded by mildly (and sub-) relativistic low-energy material. Outside the 
 highly-relativistic core, whose emission is strongly beamed, the slower material makes the dominant contribution to the flux 
 for off-axis observers due to its much wider beaming cone.
As the jet's core decelerates, its beaming cone widens and the observer sees a gradual rise in flux until the entire core becomes 
visible, at which point the flux peaks and starts to decline thereafter, gradually joining the on-axis lightcurve.
 
We demonstrate here that by using increasingly larger $\Gamma_0=20,\,40,\,60$ the initial observed time 
can be shifted to correspondingly earlier times, $t_{\rm{obs},0}=38.1,\,23.0,\,18.3\;$days, thereby replacing the sharp  rise in 
flux with a much more gradual rise. 
In GRB$\,$170817A, the shallow flux rise seen from $t_{\rm{obs,0}}\simeq10\,$days can potentially be 
reproduced for $\Gamma_0\gtrsim10^{2.5}$, which are physically plausible but computationally challenging, 
although the exact shape of the early rising lightcurve in this case is still unclear.
Nevertheless, the initially top-hat jet model has some limitations. For example, the early time afterglow lightcurve 
shows a power-law rise ($F_\nu\propto t_{\rm obs}^{0.8}$) to the peak, whereas the model lightcurve has some 
curvature. In this work we did not carry out a detailed model fit to the data to determine the goodness of fit since 
our simulations were limited to $\Gamma_0 = 60$ and could not fit observations at $t_{\rm obs}\lesssim40\,$days. 
Numerical simulations of structured jets that show a greater degree of complexity, and therefore are more realistic, 
also have larger number of model parameters, which allows them to capture the subtleties of the observed afterglow data 
more effectively.

Numerical simulations of a relativistic jet penetrating through the dynamical ejecta/neutrino-driven wind of 
BNS merger \citep{Bromberg+18,Gottlieb+18,Xie+18,Geng+19} find that the emergent jet develops a core-dominated 
angular structure similar to what we find. Moreover, our afterglow 
model fit parameters are consistent with works featuring initially structured core-dominated jets. This renders 
both scenarios practically indistinguishable from afterglow observations alone, particularly close to and after 
the peak time of the lightcurve \citep[also see, e.g.,][]{Gottlieb+19} when emission from the core starts dominating 
the observed flux, thereby validating the use of initially top-hat jet simulations as an attractive tool for afterglow 
modeling of core-dominated jets.

Both the jet's dynamics and initial angular structure outside its core, before it is decelerated by the external medium, 
affects the afterglow emission before the lightcurve peak time. From the afterglow observations alone, it might be difficult 
to disentangle their effects, however, they may be better probed by the prompt emission.
For example, in the case of GRB$\,$170817A, its highly sub-luminous and mildly soft 
prompt $\gamma$-ray emission rules out an initial top-hat jet \citep[e.g.,][]{Abbott+17-GW170817-GRB170817A,Granot+17}, favoring instead emission from sub-energetic mildly-relativistic material near our line of sight.
\acknowledgments

R.G. and J. G. are supported by the Israeli Science Foundation under grant No. 719/14.
FDC aknowledges support from the UNAM-PAPIIT grant IN117917.
We acknowledge the support from the Miztli-UNAM supercomputer (project LANCAD-UNAM-DGTIC-281) 
in which the simulations were performed.






\end{document}